%% file: Haisch.tex
\documentstyle[11pt,aaspp4]{article}
\slugcomment{Submitted to Astron. J.}
\lefthead{Haisch et al.}
\righthead{Circumstellar Disks in the IC 348 Cluster}

\begin{document}

\title{Circumstellar Disks in the IC 348 Cluster}

\author{Karl E. Haisch Jr. \altaffilmark{1,2} and Elizabeth A. Lada 
\altaffilmark{2,3}}
\affil{Dept. of Astronomy, University of Florida, 211 SSRB, Gainesville, FL  
32611}

\and

\author{Charles J. Lada \altaffilmark{2}}
\affil{Smithsonian Astrophysical Observatory, 60 Garden Street, Cambridge, 
Massachusetts 02138}

\altaffiltext{1}{NASA Florida Space Grant Fellow}

\altaffiltext{2}{Visiting Astronomer, Fred Lawrence Whipple Observatory.}

\altaffiltext{3}{Presidential Early Career Award for Scientists and Engineers Recipient.}

\begin{abstract}
We report the results of the first sensitive {\it L}-band (3.4 $\mu$m) imaging survey of the young IC 348 cluster in Perseus. In conjunction with previously acquired {\it JHK} (1.25, 1.65, 2.2 $\mu$m) observations, we use {\it L}-band data to obtain a census of the circumstellar disk population to m$_{K}$ = m$_{L}$ $\leq$ 12.0 in the central $\sim$ 110 arcmin$^{2}$ region of the cluster. An analysis of the {\it JHKL} colors of 107 sources indicates that 65\% $\pm$ 8\% of the cluster membership possesses (inner) circumstellar disks. This fraction is lower than those (86\% $\pm$ 8\% and 80\% $\pm$ 7\%) obtained from similar {\it JHKL} surveys of the younger NGC 2024 and Trapezium clusters, suggesting that the disk fraction in clusters decreases with cluster age. Sources with circumstellar disks in IC 348 have a median age of 0.9 Myr, while the diskless sources have a median age of 1.4 Myr, for a cluster distance of 320 pc. Although the difference in the median ages between the two populations is only marginally significant, our results suggest that over a timescale of $\sim$ 2 -- 3 Myr, more than a third of the disks in the IC 348 cluster disappear. Moreover, we find that at a very high confidence level, the disk fraction is a function of spectral type. All stars earlier than G appear diskless, while stars with spectral types G and later have a disk fraction ranging between 50\% -- 67\%, with the latest type stars having the higher disk fraction. This suggests that the disks around stars with spectral types G and earlier have evolved more rapidly than those with later spectral types. The {\it L}-band disk fraction for sources with similar ages in both IC 348 and Taurus is the same, within the errors, suggesting that, at least in clusters with no O stars, the disk lifetime is independent of environment.
\end{abstract}


\section{Introduction}

Multiwavelength observations of star forming regions leave little doubt that many young stellar objects (YSOs) are surrounded by circumstellar disks (\cite{ruc85}, \cite{als87}, \cite{bec90}, \cite{kh95}, \cite{hll00} (hereafter HLL00), \cite{la00}). There is much evidence to support the disk hypothesis. First, strong excess emission at infrared (e.g. \cite{ruc85}) and millimeter (e.g. \cite{bec90}) wavelengths implies that the optically thick dust must have a non-spherical distribution to preserve an optically thin line-of-sight to the star at visible wavelengths (\cite{myer87}). Second, forbidden emission-line profiles thought to be associated with mass loss in the accretion process (\cite{erm93}) are observed to be preferentially blue-shifted. The red-shifted counterpart to the preferentially blue-shifted forbidden emission lines is believed to be obscured from view by an optically thick, geometrically thin circumstellar disk. Finally, disks have been imaged directly in star-forming regions [e.g., the Orion proplyds (\cite{mco96}) and HH30 (\cite{bur96})].

Circumstellar disks are of fundamental astrophysical interest. They play a critical role in the star formation process because they are believed to be the primary channel for accreting most of the interstellar gas and dust that ultimately forms a star. More significantly perhaps, analogy with the solar system suggests that circumstellar disks are protoplanetary disks, that is, the likely formation sites of extra-solar planetary systems. Thus, determining the frequency, nature and origin of these systems is important.

While the evidence for the existence of circumstellar disks around young stars is compelling, questions regarding the formation and evolution of star/disk systems still remain. For example, what is the frequency with which disks accompany the formation of stars? What is the duration of the circumstellar disk/planet building phase of early stellar evolution? Is it a function of stellar mass or environmental (i.e., initial) conditions?

To address these very basic issues and advance our understanding of both the nature of disks and planet formation within them, we require detailed knowledge of the frequencies, lifetimes and physical properties of circumstellar disks in
populations of young stars. Most galactic field stars are believed to have originated in young embedded clusters (\cite{lada92}, \cite{pl97}, \cite{ml99}, \cite{car00}). These young clusters contain hundreds of stars covering nearly the entire range in stellar mass with similar chemical compositions in a relatively small volume at the same distance from the Sun. Therefore, the frequency of circumstellar disks presently in clusters should provide a reasonable estimate of the frequency and lifetime of protoplanetary systems within the clusters and hence an estimate of the overall likelihood of planet formation in the Galaxy.

Circumstellar disks produce infrared emission significantly in excess of that emitted by the photospheres of the young host stars, and can be readily detected by near-infrared imaging observations. Our previous near-infrared {\it L}-band (3.4 $\mu$m) and mid-infrared 10 $\mu$m studies of the young NGC 2024 cluster (HLL00, \cite{hai01}), as well as {\it L}-band imaging surveys of the Trapezium (\cite{la00}), NGC 1333 (\cite{as97}) and Taurus-Auriga (\cite{kh95}), have demonstrated that {\it L}-band observations are the optimum wavelength for detecting infrared excesses from circumstellar disks. In part, this is because disks produce sufficiently strong {\it L}-band excesses that almost all stars that have disks can be identified with {\it JHK} (1.25, 1.65, and 2.2 $\mu$m) and {\it L}-band observations, independent of inclination, disk accretion rates and inner disk hole size (\cite{kh95}). Furthermore, mid-infrared cluster observations have shown that the vast majority of the sources which have {\it L}-band excesses possess circumstellar disks (Haisch et al. 2001).

Over the past few years, we have been conducting {\it JHKL} surveys of nearby, young embedded clusters in order to establish the frequency of circumstellar disks within them. In HLL00 and Lada et al. (2000), we presented the results of our {\it L}-band surveys of NGC 2024 and the Trapezium, two very young (mean ages $\leq$ 1 Myr) clusters in Orion. These observations clearly established that the vast majority of the stars ($\geq$ 80\%) in these clusters, independent of stellar mass and stellar environment, formed with circumstellar disks. In this paper, we present the results of an {\it L}-band imaging survey of the somewhat older IC 348 cluster.

IC 348 is a relatively dense ($\rho$ $\simeq$ 105 stars pc$^{-2}$; Lada 1999) cluster located in the Perseus molecular cloud. The earliest type star in the cluster (BD+31$^{o}$643) is of spectral class B5 V. The distance to IC 348 is somewhat uncertain. Averaging parallaxes of cluster members measured by the Hipparcos satellite yields a distance of $\sim$260 pc (\cite{sch99}). This is somewhat closer than the distance of $\sim$320 pc derived from the HR diagram (Herbig 1998). The mean age of the IC 348 cluster is either 2 Myr or 3 Myr depending on the distance to the cluster (\cite{tj97}, \cite{lrll98}). Previous near-infrared {\it JHK} observations of IC 348 by Lada \& Lada (1995) have revealed approximately 350 infrared sources in the cluster. That study indicates a disk fraction of 20\% - 25\% for the YSOs in IC 348. However, for reasons discussed earlier, the {\it JHK} data do not extend to long enough wavelengths to enable either a complete or unambiguous census of the disks in the cluster.

We have used our {\it L}-band observations, in conjunction with {\it JHK} data from Lada \& Lada (1995), to obtain a census of the circumstellar disk population in the central region of the IC 348 cluster. We discuss the instrumentation, observations and data reduction in $\S$2. In $\S$3, we discuss our methods of analysis and in $\S$$\S$4 and 5 we present the results of our imaging survey. We summarize our primary results in $\S$6.

\section{Observations and Data Reduction}

The {\it L}-band (3.4 $\mu$m) observations of the IC 348 cluster were made with STELIRCAM on the Fred Lawrence Whipple Observatory (FLWO) 1.2 meter telescope on Mt. Hopkins, Arizona during the period 1998 December 03-08. STELIRCAM consists of two 256$\times$256 InSb detector arrays. Each is fed from a dichroic that separates wavelengths longer and shorter than 1.9 $\mu$m. For our survey, the dual channel camera was configured to obtain simultaneous {\it H}- and {\it L}-band observations. Three separate magnifications can be selected by rotating different cold lens assemblies into the beam. We selected a field of view of 2.5\arcmin$\times$2.5\arcmin \hspace*{0.05in}with a resolution of 0.6\arcsec \hspace*{0.05in}per pixel.

Twenty-five fields arranged in a 5$\times$5 square, covering an area of $\sim$110 arcmin$^{2}$ were observed toward the cluster. The individual images were overlapped by 30\arcsec \hspace*{0.05in}in both Right Ascension and Declination. This overlapping provided redundancy for the photometric measurements of sources located in the overlapped regions and enabled an accurate positional mosaic of the cluster to be constructed. Each individual field in the mosaic was observed in a 3$\times$3 square pattern with 12\arcsec \hspace*{0.05in}offsets or dithers between positions. The pattern was observed such that the telescope was not offset in the same direction twice. Once the 9 dithered positions had been observed, the telescope was offset by 5\arcsec \hspace*{0.05in}in a random direction and the pattern repeated. The integration time at each dither position was 10 seconds (0.1 second x 100 coadditions), which ensured that the counts acquired per exposure were at least half of the full-well capacity of the array. The total integration time for each field of 3 minutes.

All data were reduced using the Image Reduction and Analysis Facility (IRAF). An average dark frame was constructed from the dark frames taken at the beginning and end of each night's observations. This dark frame was subtracted from all target observations to yield dark subtracted images. The data were then linearized. Linearization is typically the first step in the reduction procedure. However, the data were obtained using a single readout mode, and the bias applied to the array is still present in each frame. Thus, in order to properly linearize the data, we must first remove this bias, hence the initial dark subtraction. Sky frames were individually made for each observation by median combining the nearest nine dark subtracted frames in time to the target observation. Each sky frame was checked to confirm that all stars had been removed by this process. The sky frames were then normalized to produce flat fields for each target frame. All dark subtracted target frames were then processed by subtracting the appropriate sky frame and dividing by the flat field. Finally all target frames for a given position in the cluster were registered and combined to produce the final image of each field.

The {\it L}-band data were combined with {\it JHK} observations using the Simultaneous Quad Infrared Imaging Device (SQIID) on the 1.3 meter telescope at Kitt Peak National Observatory. Details of the {\it JHK} data can be found in Lada \& Lada (1995).

\section{Analysis}

\subsection{Source Extraction and Photometry}

Infrared sources were identified using the DAOFIND routine (\cite{stet87}) within IRAF. The full width at half maximum (FWHM) of the point spread function (PSF) ranged from 2.3 to 2.8 pixels. DAOFIND was run on each individual image using a FWHM of the PSF between 2.0 and 2.5 pixels and a single pixel finding threshold equal to 3 times the mean noise of each image. Each frame was individually inspected and the DAOFIND coordinate files were edited to remove bad pixels and any objects misidentified as stars, as well as add any missed stars to the list. Aperture photometry was then performed using the routine PHOT within IRAF. An aperture of 4 pixels in radius was used for all photometry. Aperture diameters were selected such that they were at least twice the FWHM of the PSF of the stars. The sky values around each source were determined from the mode of intensities in an annulus with inner and outer radii of 10 and 20 pixels respectively. Our choice of aperture sizes insured that the flux from at least 90\% of the stars was not contaminated by the flux from neighboring stars, however they are not large enough to include all the flux from a given source. In order to account for this missing flux, aperture photometry was performed on all bright, isolated sources using the same aperture used for the photometry of the standard stars, 10 pixels for each dataset. Fluxes in both the large and small apertures were compared and the instrumental magnitudes for all sources were corrected to account for the missing flux.

Photometric calibration was accomplished using the list of standard stars of Elias et al. (1982). The standards were observed on the same nights and through the same range of airmasses as the cluster and control fields. Zero points and extinction coefficients were established for each night. Due to the extensive spatial overlapping of the cluster images, a number of cluster sources were observed at least twice. We compared the {\it L}-band magnitudes of the sixty duplicate stars identified in the overlap regions. For all stars brighter than the completeness limit of our survey, a comparison of the photometry of the duplicate stars agreed to within 10\%.

\subsection{Completeness Limit}

The completeness limit of the observations was determined by adding artificial stars to each of the images and counting the number of sources recovered by DAOFIND. Artificial stars were added at random positions to each image in eight separate half magnitude bins with each bin containing one hundred stars. The eight bins covered a magnitude range from 10.0 to 14.0. The artificial stars were examined to insure that they had the same FWHM of the PSF as the real sources in the image.  Aperture photometry was performed on all sources to confirm that the assigned magnitudes of the added sources agreed with those returned by PHOT. All photometry agreed to within 0.1 magnitudes. DAOFIND and PHOT were then run and the number of identified artificial sources within each half magnitude bin was tallied. This process was repeated 20 times. We estimate that the identification of sources in our STELIRCAM survey is 90\% complete to {\it m}$_{L}$ = 12.0. The {\it JHK} data were estimated to be 90\% complete to m$_{J}$ = 16.25, m$_{H}$ = 15.25, and m$_{K}$ = 14.25 (\cite{ll95}).

\subsection{Astrometry}

Absolute $\alpha$ and $\delta$ positions were determined for all sources in our survey. A centering algorithm in APPHOT was used to obtain center positions for the sources in pixels. Many sources were present in the overlap regions of our IC 348 fields. Duplicate stars on adjacent frames were used to register the relative pixel positions of the individual frames in each survey onto a single positional grid. Pixel positions were converted to equitorial coordinates using the position of BD+31$^{o}$643 as a reference and a plate scale of 0.6\arcsec/pixel. A list of all objects for which we have complete {\it JHKL} photometry is given in Table~\ref{phottable}. Also listed in Table~\ref{phottable} are the RA (2000) and Dec (2000) coordinates and the near-infrared colors for each source ({\it JHK} magnitudes were taken from \cite{ll95}). The average difference in the positions between our survey and those from Lada \& Lada (1995) is less than $\sim$ 1\arcsec. We are therefore quite certain of the matches between the {\it JHK} and {\it L}-band data. Finally, we have compared our source coordinates with those in the 2MASS database for IC 348. The coordinates for all sources agree to within an average of 1\arcsec.

\section{Results}

\subsection{Color-Color Diagrams: Infrared Excess Fractions}

In Figure \ref{jhklcols}, we present the {\it JHK} and {\it JHKL} color-color diagrams for the IC 348 cluster. In the diagrams, we have included only those 107 sources from our STELIRCAM survey which have m$_{K}$ = m$_{L}$ $\leq$ 12.0. By selecting sources in this way, our survey only includes those sources whose photospheres are detectable at both {\it K}- and {\it L}-bands. This insures a meaningful determination of the fraction of the sources having an infrared excess, and thus the circumstellar disk fraction (see $\S$5). In the diagrams, we plot the locus of points corresponding to the unreddened main sequence as a solid line and the locus of positions of giant stars as a heavy dashed line (Bessell \& Brett 1988). The two leftmost parallel dashed lines define the reddening band for main sequence stars and are parallel to the reddening vector. Crosses are placed along these lines at intervals coresponding to 5 magnitudes of visual extinction. The classical T Tauri star (CTTS) locus is plotted as a dot-dashed line (\cite{mch97}). The reddening law of Cohen et al. (1981), derived in the CIT system and having slopes in the {\it JHK} and {\it JHKL} color-color diagrams of 1.692 and 2.750 respectively, has been adopted.

A significant fraction of the cluster sources fall outside and to the right of 
the reddening lines in the infrared excess region of the {\it JHK} and {\it JHKL} color-color diagrams. Indeed, a total of 19/107 (18\% $\pm$ 4\%) and 59/107 (55\% $\pm$ 7\%) of the sources lie in the infrared excess region of the {\it JHK} and {\it JHKL} color-color diagrams respectively. We must correct these fractions for field star contamination. The foreground/background contamination was determined by first counting the total number of stars observed in the control fields from the Lada \& Lada (1995) {\it JHK} survey of IC 348 to m$_{K}$ = 12. Scaling this number to the cluster area observed and accounting for the average extinction in IC 348 (A$_{K}$ $\simeq$ 0.5 magnitudes; \cite{ll95}), we find a field star contamination of 16 stars. Applying this correction to the infrared excess fractions, we estimate that 21\% $\pm$ 5\% and 65\% $\pm$ 8\% of the sources in the IC 348 cluster have colors exhibiting {\it JHK} and {\it JHKL} infrared excess emission respectively.

As discussed in HLL00 and Lada et al. (2000), the derived fraction of excess sources is sensitive to the location of the boundary of the reddening band. This depends on the adopted intrinsic colors of the latest spectral type stars. At our m$_{K}$ = m$_{L}$ = 12.0 completeness limit, we are sensitive to stars with spectral classes down to M5 (\cite{lrll98}), which corresponds to a mass of 0.19 M$_{\odot}$ for a mean cluster age of 2 -- 3 Myr. Furthermore, this also corresponds to the boundary adopted in our previous {\it L}-band cluster studies (HLL00, Lada et al. 2000). The {\it JHKL} excess fraction for IC 348 has therefore been calculated assuming the boundary of the reddening band in the {\it JHKL} diagram passes through M5 colors from Bessell \& Brett (1988). In the {\it JHK} diagram, however, we use A0 colors from Bessell \& Brett (1988) since if one extends the right-hand reddening line from an M5 spectral class toward bluer colors, some of the earlier type main-sequence stars fall in the infrared excess region of the diagram. This reddening line extends through main-sequence colors that are somewhat later than M5, however we consider the {\it JHKL} fraction as derived above to be a lower limit on the true excess fraction.

The frequency of near-IR excess sources determined from color-color diagrams also depends on the adopted reddening law and to a lesser extent on the photometric system used to plot the positions of the main sequence stars and the reddening bands. We have calculated the {\it JHK} and {\it JHKL} infrared excess fractions for two other reddening laws obtained in different photometric systems (\cite{kor83}; \cite{rl85}), and find {\it JHK} excess fractions of 21\% $\pm$ 5\% and 20\% $\pm$ 5\%, and {\it JHKL} excess fractions of 65\% $\pm$ 8\% and 60\% $\pm$ 8\%, respectively. These fractions are within the quoted (statistical) error of our determination using a Cohen et al. (1981) reddening law.

Finally, we note that photometry at {\it L}-band can be affected for heavily obscured sources if the source spectrum contains a 3.08 $\mu$m ice absorption feature. The effects of this feature on near-infrared photometry have been investigated by a number of authors (e.g. \cite{whit88}; \cite{snty90}; \cite{arl94}; \cite{as97}). The ice feature suppresses the flux in the {\it L} band, thereby making the {\it K} -- {\it L} color too blue. In the IC 348 cluster, the majority of the sources are extincted by A$_{V}$ $\leq$ 5 magnitudes (\cite{ll95}). Sato et al. (1990) have defined an empirical relationship between the optical depth of the ice feature, $\tau$$_{ice}$, and A$_{V}$ for the Taurus region, namely: $\tau$$_{ice}$ = 0.093(A$_{V}$ -- A$_{vc}$), where A$_{vc}$ = 1.6. Under the assumption that this equation is valid in IC 348, we find that most of the stars in the cluster have $\tau$$_{ice}$ $\leq$ 0.32. Aspin \& Sandell (1997) have calculated the narrow-band {\it L} magnitude correction due to ice absorption in the NGC 1333 cluster using the Sato et al. (1990) $\tau$$_{ice}$-A$_{V}$ relation. For a source with $\tau$$_{ice}$ = 0.32, the narrow-band {\it L} correction is $\sim$ 0.1 magnitude. The correction to our IC 348 broad-band {\it L} photometry will be even smaller due to the increased continuum flux from the stellar source, which would dilute the line strength of the ice feature relative to that observed in a narrow-band {\it L} filter. We therefore do not expect the ice feature to produce a large effect on our {\it L}-band photometry. Nevertheless, we consider the excess fractions quoted above to be lower limits, and the true {\it JHKL} excess fraction may be somewhat higher if ice absorption is present.

\subsection{H$\alpha$ Emission Stars and {\it JHKL} Colors}

In Figure~\ref{jhklha}, we present the {\it JHKL} color-color diagram showing the location of those 60 sources in our survey which had unambiguous counterparts in the H$\alpha$ survey of Herbig (1998). Filled circles represent those sources with positive detections, that is sources which have equivalent widths EW(H$\alpha$) $\geq$ 2 \AA, and open circles denote sources with EW(H$\alpha$) $<$ 2 \AA. Among the H$\alpha$ detections, 24/32 (75\% $\pm$ 15\%) lie in the infrared excess region of the {\it JHKL} color-color diagram. Of the sources which were not detected in H$\alpha$, 7/28 (25\%) lie in the infrared excess region of the color-color diagram, similar to the number (8/32; 25\%) of sources with H$\alpha$ detections which lie in the reddening band.

We divide the H$\alpha$ detections into classical T Tauri stars (CTTS; EW(H$\alpha$) $\geq$ 10 \AA) and weak-line T Tauri stars (WTTS; EW(H$\alpha$) $<$ 10 \AA) (\cite{am89}, \cite{men99} and references therein) and show their locations in the {\it JHKL} color-color diagram in Figure~\ref{jhkltts}. CTTSs are denoted by a filled circle, while WTTSs are shown with an open circle. In the diagram we have only plotted those WTTSs with 2 $<$ EW(H$\alpha$) $<$ 10 \AA. We adopt this range of H$\alpha$ equivalent widths for the WTTSs in our subsequent discussion. Among the H$\alpha$ detections, we identify 17 WTTS and 15 CTTS. All but one (14/15; 93\%) of the CTTSs lie in the infrared excess region of the color-color diagram. Among the WTTSs, 10/17 (59\%) have {\it JHKL} colors indicative of an infrared excess.

\subsection{Effects of Source Variability}

Source variability could affect our derived disk fraction since the {\it L}-band observations were taken roughly seven years after the {\it JHK} data. The canonical view for the source of the variability is that in WTTSs, the variability is due to large, cool spots on the stellar surface, while in CTTSs there is an additional component caused by accretion hot spots (which have shorter timescales). In order to investigate the overall effects of variability on our computed disk fraction, we compared the SQIID {\it H}-band magnitudes of the 107 sources in our survey with the STELIRCAM {\it H}-band magnitudes acquired in conjunction with our {\it L}-band data. Sixteen ($\sim$15\%) had fluxes which differed by more than 0.15 magnitudes. These sources are labelled in Table~\ref{phottable}. In addition, we were able to match 104 of our sources (which includes the 16 sources just discussed) with objects in the Herbst, Maley, \& Williams (2000) study of variability in IC 348. Ten ($\sim$ 10\%) of these were identified as likely variable stars by Herbst, Maley, \& Williams (2000). Thus, variability could have significantly affected the photometry of 10 -- 15\% of the sources in the cluster. However, variability would move equal numbers of sources to the right and to the left in the color-color diagram, and therefore one would lose, as well as gain, excess sources. Therefore, on average, this would have only a small effect on the excess fraction.  Finally, variability has also been shown to have a negligible effect on the infrared excesses found in other star forming regions (e.g. \cite{bar97}, HLL00, Lada et al. 2000). We therefore find it very unlikely that source variability is a major factor in producing infrared excesses, and thus has little effect on our derived disk fraction.

As noted in the previous section, eight of the sources detected in H$\alpha$ fall in the reddening band of the {\it JHKL} color-color diagram, and seven H$\alpha$ non-detections lie in the infrared excess region. Some of the H$\alpha$ detections may have disks, while the non-detections likely do not have disks (see $\S$5). Source variability may be responsible for the location of these sources in the {\it JHKL} diagram. A comparison of the {\it H} band STELIRCAM and SQIID magnitudes, in conjunction with the variability measure $\sigma$$_{var}$ of Herbst, Maley, \& Williams (2000), reveals that 3 of the H$\alpha$ detections and 4 of the H$\alpha$ non-detections have {\it H} band magnitudes which differ by $\geq$ 0.1 magnitude or $\sigma$$_{var}$ $>$ 0.15 magnitude. Hence, source variability can account for the locations of 7 of the sources in the {\it JHKL} color-color diagram. Photometric error could be responsible for the location of the 3 remaining H$\alpha$ non-detections with infrared excesses since they are near the reddening band. The location of the remaining 5 sources with H$\alpha$ detections in the reddening band is likely the result of chromospheric activity rather than a disk as discussed in the next section.

\section{Discussion}

Using our {\it L}-band photometry in combination with {\it JHK} observations obtained by Lada \& Lada (1995) has allowed us to determine a more robust estimate of the circumstellar disk fraction in the IC 348 cluster than has previously been possible. The infrared excess fraction determined from the {\it JHKL} color-color diagram for the IC 348 cluster is 65\% $\pm$ 8\%. Our previous {\it L}-band and mid-infrared studies of young clusters have demonstrated that the most likely source of the {\it JHKL} infrared excesses is an inner circumstellar disk (HLL00, Lada et al. 2000, Haisch et al. 2001). That the {\it JHKL} disk fraction in IC 348 is much larger than that inferred from {\it JHK} colors (20\% - 25\%; \cite{ll95}, \cite{lrll98}) is consistent with the predictions of circumstellar disk models (\cite{la92}, \cite{mch97}) and confirms similar results for the NGC 2024 and Trapezium clusters (HLL00, Lada et al. 2000).

Further evidence of a significant circumstellar disk population in the IC 348 cluster comes from the fraction of the H$\alpha$ stars (75\% $\pm$ 15\%) which have infrared excesses in the {\it JHKL} color-color diagram. The majority (14/15; 93\%) of the stars identified as CTTSs have {\it JHKL} colors indicative of circumstellar disks. It is interesting, however, that a large fraction (10/17; 59\%) of the sources identified as WTTSs (c.f. those with 2 $<$ EW(H$\alpha$) $<$ 10 \AA) display infrared excesses. The widely held view is that CTTSs are surrounded by accretion disks while WTTSs are not (e.g. \cite{ob95}; \cite{men99}). Nevertheless, the infrared excesses observed in some of the WTTSs suggests the presence of circumstellar disks surrounding these objects. 

Walter \& Barry (1991) have shown that, in low mass WTTSs in Taurus-Auriga, chromospheric activity produces H$\alpha$ emission with equivalent widths of $\sim$ 2 -- 10 \AA. However, these authors also show that, in low mass PMS stars whose H$\alpha$ emission is attributed to circumstellar disks, the equivalent widths can vary widely, from $\sim$ 2 \AA \hspace*{0.05in}to greater than 10 \AA, and can thus have comparable H$\alpha$ equivalent widths to low mass WTTSs whose H$\alpha$ emission is produced by chromospheric activity. Therefore, while low mass PMS stars with H$\alpha$ equivalent widths $\geq$ 10 \AA \hspace*{0.05in}likely have disks, using H$\alpha$ as an indicator of the presence or absence of a circumstellar disk for low mass PMS stars with 2 $<$ EW(H$\alpha$) $<$ 10 \AA \hspace*{0.05in}is ambiguous (stars with EW(H$\alpha$) $<$ 2 \AA \hspace*{0,05in}probably do not have disks). A similar conclusion has also been reached by Mart\'{i}n (1997).

Our previous studies have shown that {\it K} -- {\it L} colors will identify all disks in a population of young stars (HLL00; Lada et al. 2000; Haisch et al. 2001). Thus, {\it L}-band observations remove the ambiguity which is present in using H$\alpha$ as a disk indicator for stars with 2 $<$ EW(H$\alpha$) $<$ 10 \AA, and are therefore superior to H$\alpha$ in identifying the presence of disks around young stars. The presence of WTTSs in the infrared excess region of the {\it JHKL} color-color diagram demonstrates that stars with 2 $<$ EW(H$\alpha$) $<$ 10 \AA \hspace*{0.05in}can have circumstellar disks. The H$\alpha$ emission in the WTTSs which lie in the reddening band is likely due to chromospheric activity in diskless sources. Some WTTSs in Taurus-Auriga also lie in the infrared excess region of the {\it JHKL} color-color diagram. In Figure \ref{jhklttstau}, we show the {\it JHKL} color-color diagram for CTTSs and WTTSs in Taurus-Auriga to the equivalent completeness limit of our IC 348 observations (ie. corrected for distance). The data in Figure \ref{jhklttstau} were taken from Strom et al. (1989) and Kenyon \& Hartmann (1995). Filled circles represent CTTSs while open circles denote WTTSs (with 2 $<$ EW(H$\alpha$) $<$ 10 \AA). As in IC 348, some of the WTTSs have infrared colors signifying the presence of a disk. Taken together, these results confirm both earlier indications that WTTS stars are not necessarily diskless (e.g. \cite{bran00}), and that {\it K} -- {\it L} colors will identify all disks in a population of young stars, and hence is superior to H$\alpha$ in identifying the presence of disks. We note here that the WTTSs with disks in both IC 348 and Taurus-Auriga do not show infrared excesses in {\it JHK} color-color diagrams. Thus, while CTTSs and WTTSs both appear to have disks, the magnitude of the accretion taking place in the disks may differ.

The disk fraction (65\% $\pm$ 8\%) in the IC 348 cluster is lower than the disk fractions (86\% $\pm$ 8\% and 80\% $\pm$ 7\%) in the NGC 2024 and Trapezium clusters respectively (HLL00, \cite{la00}). The mean age of the stars in the core of the IC 348 cluster is either 2 Myr or 3 Myr depending on the cluster distance (320 pc vs. 260 pc), while the NGC 2024 and Trapezium clusters are younger (mean ages of 0.3 Myr and 0.8 Myr respectively). Taken together, these results clearly indicate a decrease in the circumstellar disk fraction with cluster age.

We note that the {\it L}-band emission from the disks is produced very close ($\leq$ 0.1 AU) to the stellar surface. However, it is likely that the decrease in the disk fraction that we observe in IC 348 applies to the outer ($\geq$ 1 AU) disks as well as the inner disks. Evidence for this derives from millimeter studies (which probe the outer disks) of WTTSs (those stars with 0 $<$ EW(H$\alpha$) $<$ 10 \AA) in Taurus-Auriga, a region of similar mean age to IC 348. A total of 44 WTTSs are common to both the {\it L}-band (\cite{kh95}) and millimeter continuum (\cite{bec90}, \cite{ob95}) studies of Taurus-Auriga. Eleven of the 44 sources (25\% $\pm$ 8\%) exhibit {\it L}-band excess emission, similar to the 10/44 (23\% $\pm$ 7\%) which have detections at 1.3 mm. In addition, only 3 of the 34 (9\% $\pm$ 5\%) sources not detected at millimeter wavelengths exhibited {\it L}-band excess emission. Similarly, only 2/33 (6\% $\pm$ 4\%) of the sources without {\it L}-band excesses were detected at 1.3 mm. These results indicate that stars lacking a signature of an inner disk also lack an outer disk. Finally, we find an (inner) disk fraction of 17/45 (38\% $\pm$ 9\%) among the WTTSs (in this case, those stars with 0 $<$ EW(H$\alpha$) $<$ 10 \AA) in IC 348, similar to the fraction of WTTS with millimeter continuum detections in Taurus-Auriga.  Thus, it is likely that both our derived disk census and disk lifetime for IC 348 apply to the outer as well as the inner circumstellar disk region.

Ages for the individual sources in the IC 348 cluster were determined by Kevin Luhman for cluster distances of 260 pc and 320 pc using the models of D'Antona \& Mazzitelli (1997) and luminosities and dwarf temperatures from Luhman et al. (1998) and Luhman (1999). Depending on the adopted distance, sources with circumstellar disks have a median age of 0.9 or 1.7 ($\pm$ 0.2) Myr, while the diskless sources have a median age of 1.4 or 2.2 ($\pm$ 0.4) Myr, where the older age corresponds to the closer distance. A Kolmogorov-Smirnov (K-S) test indicates a probability of {\it p$_{K-S}$} $\simeq$ 0.26 that the ages of the disk and diskless stars were drawn randomly from the same parent population. While this is only at the 1$\sigma$ level of significance, this still suggests that the sources without disks may be older than those with disks. Furthermore, our results suggest that the timescale for more than a third of the disks in the IC 348 cluster to disappear is $\sim$ 2 -- 3 Myr.

Strom et al. (1989) and Strom, Edwards, \& Skrutskie (1993) have investigated the disk evolution timescales in the Taurus-Auriga star forming region, which has a similar mean age to IC 348. Of the 83 sources in these studies with complete {\it JHKL} photometry, 44/78 (56\% $\pm$ 8\%) showed infrared excesses indicative of disks. This is the same, to within errors, as the fraction in IC 348 (65\% $\pm$ 8\%). IC 348 represents an environment with a much higher stellar density relative to Taurus-Auriga, where star formation occurs in relative isolation. Our previous work has shown that high stellar density does not appear to inhibit the {\it formation} of circumstellar disks (HLL00, Lada et al. 2000). A comparison of the disk fractions and source ages in IC 348 and Taurus suggests that, at least in clusters with no O stars, the {\it disk lifetime} is also independent of the star forming environment.

Published spectral types are available for 80 of the sources in our sample (\cite{lrll98}). Table~\ref{spectable} lists the infrared excess fractions (not corrected for foreground/background contamination) as a function of spectral type (counting from the M5 boundary). Significantly, the circumstellar disk fraction appears to depend on the spectral type, or mass, of the stars in the cluster. None of the stars with spectral types earlier than G appear to have disks. In contrast, $\geq$ 50\% of all stars with spectral types of G and later appear to have disks, with the largest disk fraction (67\%) found around the M type stars. A K-S test indicates a very low probability of {\it p$_{K-S}$} $\simeq$ 4 $\times$ 10$^{-8}$ that the disk fractions for the high and low mass stars were drawn randomly from the same parent population. In the younger Trapezium cluster, circumstellar disks appear to form with equally high frequency ($\sim$ 80\%) around all stars with spectral types F and later, with a lower, but nevertheless significant, frequency (42\%) around stars with earlier OBA spectral types (\cite{la00}). These results suggest that the dissipation rate of circumstellar disks is a function of both time and stellar mass, such that disks around stars with spectral types earlier than G have shorter lifetimes compared to disks surrounding stars of later spectral types.

%Spectral types exist for all of the WTTS and 11 of the CTTS stars in our %sample. We can combine these spectral types and {\it K} -- {\it L} colors from %our study with the variability measure $\sigma$$_{var}$ from HMW00 to %reinvestigate the relationship between infrared excess and variability as %studied by the aforementioned authors. We first compute the intrinsic color %excesses for each star using the definition from Meyer, Calvet, \& Hillenbrand %(1997), namely
%
%\begin{equation}
%E({\it K} -- {\it L})_{o} = ({\it K} -- {\it L}) -- ({\it K} -- {\it L})_{*} -- %0.004 \times A_{v}
%\end{equation}
%
%\noindent where the first term is the observed color of the star, the second %term is the intrinsic stellar color based on its spectral type and the third %term is due to extinction. We calculated the second term using the intrinsic %photospheric colors from Bessell \& Brett (1988). Extinctions were calculated %using the Bessell \& Brett colors and dereddening the stars either to the main %sequence (for sources within the reddening band of the {\it JHKL} color-color %diagram) or the CTTS locus (for stars in the infrared excess region of the %diagram).
%
%In Figure \ref{ic348var}, we show the degree of variability, $\sigma$$_{var}$, %versus the intrinsic color excesses for each star. There appears to be no %correlation between variability and infrared excess, consistent with the %results of HMW00 which were based on {\it I} -- {\it K} colors. Mention star %73.....

\section{Summary and Conclusions}

We have obtained the first sensitive {\it L}-band survey of a $\sim$110 arcmin$^{2}$ region of the IC 348 cluster in Perseus. These data, in conjunction with previously obtained {\it JHK} observations, were used to produce a relatively unambiguous census of the infrared excess/circumstellar disk population in the cluster. The primary conclusions derived from our study are summarized as follows:

1. We detected 107 sources with {\it K} = {\it L} $\leq$ 12.0 in our survey region. An analysis of the {\it JHKL} colors indicates a circumstellar disk fraction of 65\% $\pm$ 8\%. This fraction is lower than those (86\% $\pm$ 8\% and 80\% $\pm$ 7\%) obtained for the younger NGC 2024 and Trapezium clusters, suggesting that the circumstellar disk fraction decreases with cluster age.

2. At a very high confidence level, we find that the lifetime of circumstellar disks is dependent on spectral type or stellar mass, with disks around massive stars dissipating more quickly than those around less massive stars. None of the sources with spectral types earlier than G have disks. For stars with later spectral types, $\geq$ 50\% have disks, with the largest disk fraction (67\%) found around the M stars. In contrast, the disk fraction in the Trapezium cluster is independent of spectral type for all stars with spectral types F and later, with a lower, but nevertheless significant, disk frequency around stars with earlier spectral types. 

3. Sources with circumstellar disks in IC 348 have a median age of 0.9 (1.7) Myr, while the diskless sources have a median age of 1.4 (2.2) Myr, for cluster distances of 320 (260) pc. A K-S test indicates only a marginal significance in the difference in median ages between the two populations, nonetheless our results suggest that the timescale for more than a third of the disks in the IC 348 cluster to disappear is $\sim$ 2 -- 3 Myr. However, {\it all} stars with spectral types of F and earlier have lost their disks on the same timescale. While the {\it L}-band emission arises primarily in the inner disk regions, comparisions of {\it L}-band and millimeter continuum data for WTTSs in Taurus-Auriga, in conjunction with our IC 348 WTTS {\it L}-band results, suggest that the inner and outer disks disappear on similar timescales.

4. The disk fraction for sources in IC 348 and Taurus-Auriga (which have similar mean ages) is the same, to within the errors. IC 348 represents an environment with a much higher stellar density relative to Taurus-Auriga, where star formation occurs in relative isolation, suggesting that, at least in clusters with no O stars, disks are not destroyed on shorter timescales in dense cluster environments.

5. All but one of the CTTSs identified in our sample have {\it JHKL} colors indicative of disk emission. Interestingly, a large fraction of the WTTSs (those with 2 $<$ EW(H$\alpha$) $<$ 10 \AA) also display infrared excesses indicative of circumstellar disks, confirming both earlier indications that WTTS stars are not necessarily diskless, and that {\it K} -- {\it L} colors will identify all disks in a population of young stars, and hence is superior to H$\alpha$ in identifying the presence of disks. The WTTSs with disks do not show infrared excesses in the {\it JHK} color-color diagram, suggesting that while CTTSs and WTTSs both appear to have disks, the magnitude of the accretion taking place in the disks may differ.

\acknowledgements

We are very grateful to Kevin Luhman for providing ages for many of the cluster sources based on two different distances. We also thank Eric Tollestrup for the development, use and support of STELIRCAM. K. E. H. gratefully acknowledges support from a NASA Florida Space Grant Fellowship and an ISO grant through JPL \#961604. E. A. L. acknowledges support from a Research Corporation Innovation Award and a Presidential Early Career Award for Scientists and Engineers (NSF AST 9733367) to the University of Florida. We also acknowledge support from an ADP (WIRE) grant NAG 5-6751.

\newpage
\input{figcaptions}
\input{tables}
\input{figurelist}

\end{document}

%% file: figcaptions.tex
\figcaption[Haisch.fig1.eps]
{{\it JHK} ({\it left}) and {\it JHKL} ({\it right}) color-color diagrams for the IC 348 cluster. Only those sources with m$_{K}$ = m$_{L}$ $\leq$ 12.0 and {\it JHKL} photometric errors less than 10\% are plotted. In the diagram, the locus of points corresponding to the unreddened main sequence is plotted as a solid line, the locus of positions of giant stars is shown as a heavy dashed line and the CTTS locus as a dot-dashed line. The two leftmost dashed lines define the reddening band for main sequence stars and are parallel to the reddening vector. Crosses are placed along these lines at intervals corresponding to 5 magnitudes of visual extinction. The rightmost dashed line is parallel to the reddening band.
\label{jhklcols}
}

\figcaption[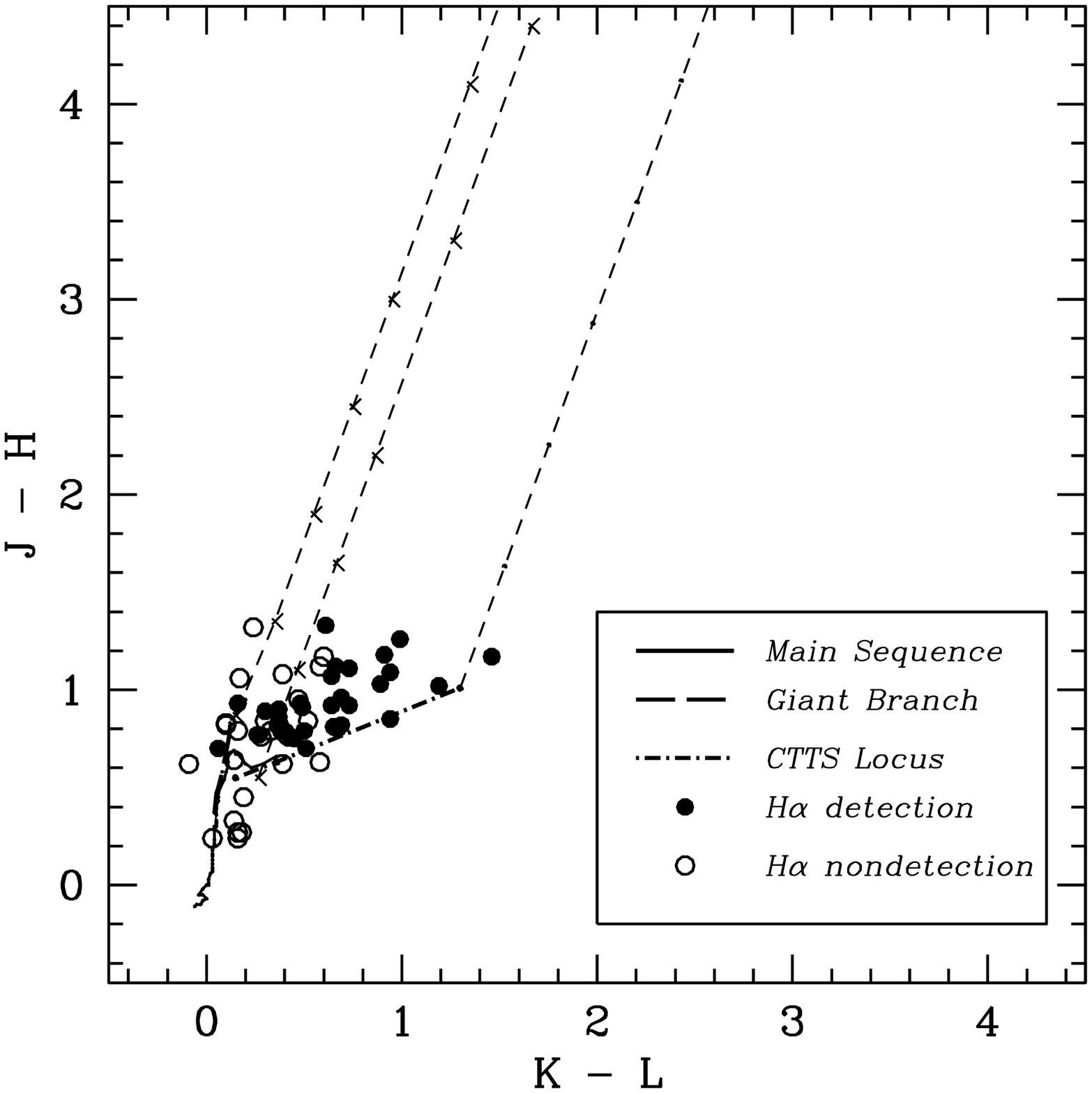]
{{\it JHKL} color-color diagram for IC 348 showing the locations of sources with H$\alpha$ detections from Herbig (1998). Sources with H$\alpha$ detections are shown as a solid circle, while stars not detected in H$\alpha$ are shown as open circles.
\label{jhklha}
}

\figcaption[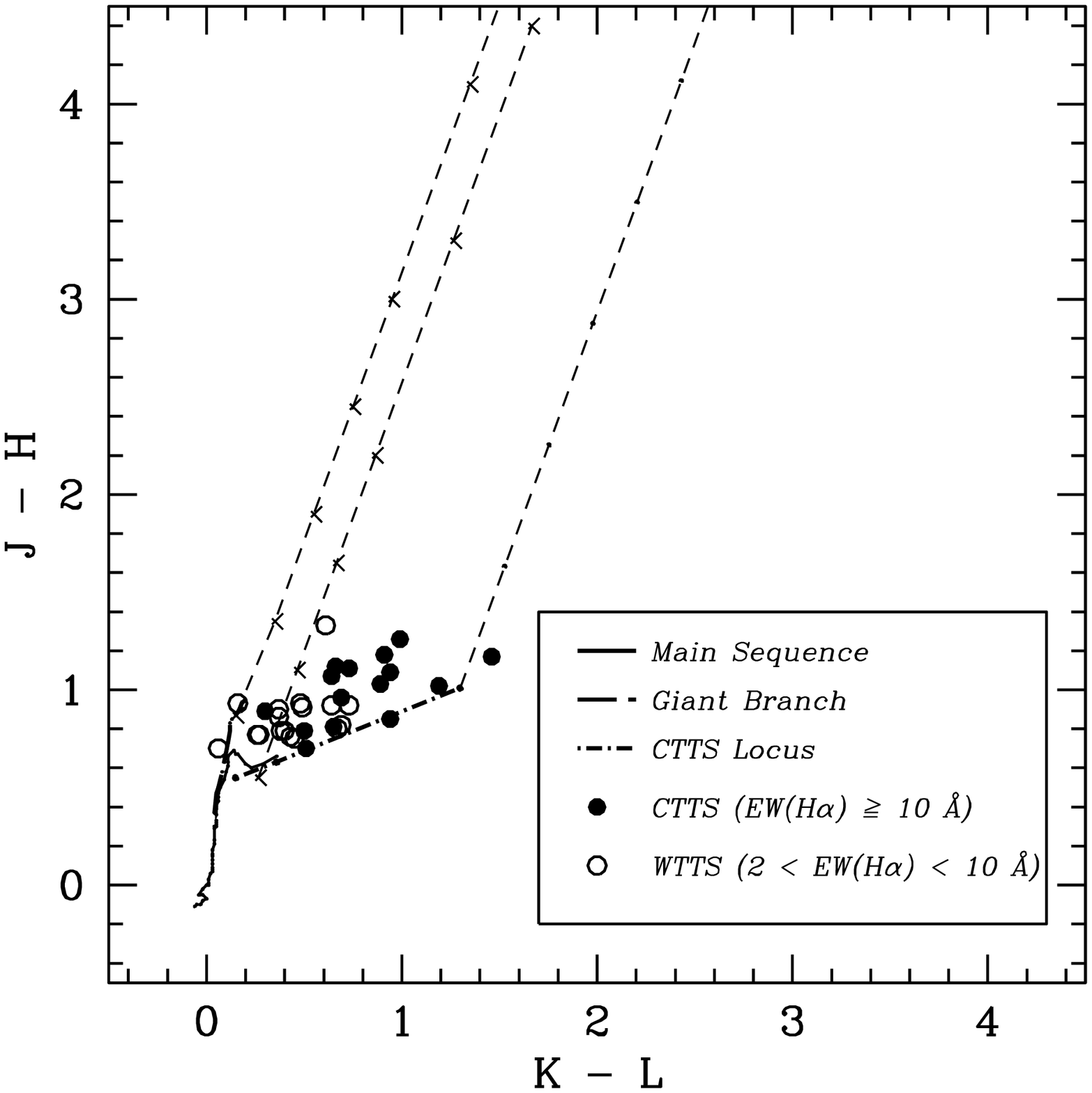]
{{\it JHKL} color-color diagram for IC 348 showing the locations of CTTS and WTTS sources. CTTS are shown as a solid circle, while WTTS are shown as open circles. Note that we only plot those WTTSs with 2 $<$ EW(H$\alpha$) $<$ 10 \AA.
\label{jhkltts}
}

\figcaption[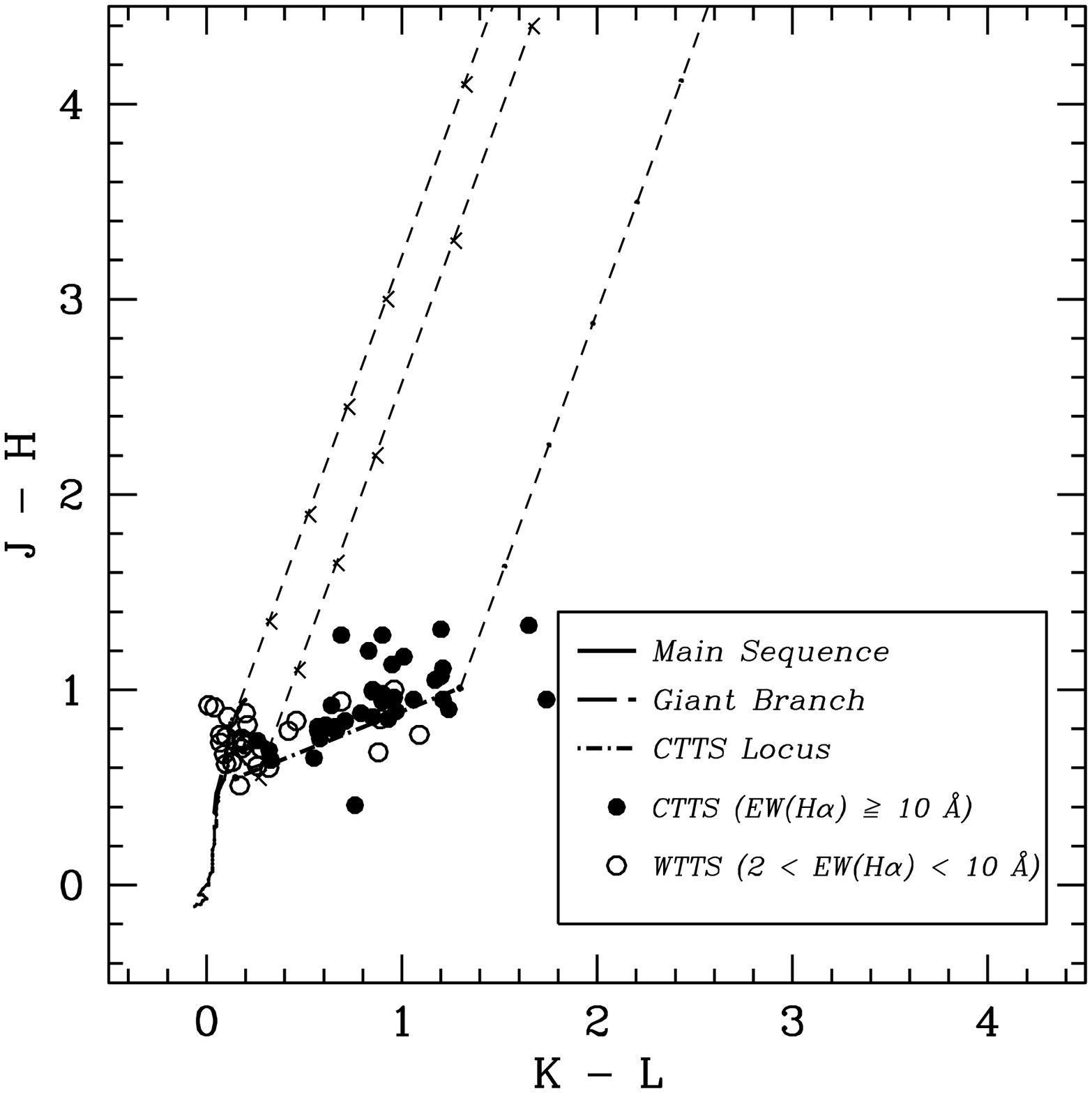]
{{\it JHKL} color-color diagram for Taurus-Auriga showing the locations of CTTS and WTTS sources. CTTS are shown as a solid circle, while WTTS are shown as open circles. The data were taken from Strom et al. (1989) and Kenyon \& Hartmann (1995).
\label{jhklttstau}
}

%% file: tables.tex
\clearpage
\input{Haisch.tab1.tex}
\clearpage
\input{Haisch.tab2.tex}

\clearpage

%% file: Haisch.tab1.tex
\begin{deluxetable}{cccccccccc}
\footnotesize
\tablecaption{{\it JHKL} Photometry of IC 348 Sources \label{phottable}}
\tablewidth{0pt}
\tablehead{Star ID\tablenotemark{1} & RA (2000) & Dec (2000) & J\tablenotemark{2} &
H\tablenotemark{2} & K\tablenotemark{2} & L & (J-H) & (H-K) & (K-L)}
\startdata
   1 & 3 44 34.15 & 32 09 44.58 &  7.74 &  7.63 &  6.97 &  6.44  &   0.11 & 0.66 & 0.53\nl
   2 & 3 44 35.36 & 32 10 02.71 &  8.28 &  8.01 &  7.45 &  6.87  &   0.27 & 0.56 & 0.58\nl
   3 & 3 44 31.07 & 32 06 20.26 &  8.55 &  8.28 &  7.93 &  7.77  &   0.27 & 0.35 & 0.16\nl
   4$^{*}$ & 3 44 25.92 & 32 04 29.97 & 10.15 &  8.97 &  7.95 &  7.04  &   1.18 & 1.02 & 0.91\nl
   5$^{*}$ & 3 44 36.89 & 32 06 43.11 &  9.22 &  8.59 &  8.21 &  7.84  &   0.63 & 0.38 & 0.37\nl
   6 & 3 44 08.35 & 32 07 17.23 &  8.91 &  8.78 &  8.66 &  8.51  &   0.13 & 0.12 & 0.15\nl
   7 & 3 44 09.05 & 32 07 09.99 &  8.93 &  8.77 &  8.68 &  8.59  &   0.16 & 0.09 & 0.09\nl
   8$^{*}$ & 3 44 39.15 & 32 09 15.98 &  9.94 &  9.15 &  8.75 &  8.42  &   0.79 & 0.40 & 0.33\nl
   9 & 3 44 24.64 & 32 10 14.05 &  9.30 &  8.97 &  8.82 &  8.68  &   0.33 & 0.15 & 0.14\nl
  10 & 3 44 32.09 & 32 11 42.90 &  9.95 &  9.32 &  8.90 &  8.32  &   0.63 & 0.42 & 0.58\nl
  11 & 3 44 55.04 & 32 12 08.52 & 10.28 &  9.35 &  8.94 &  8.59  &   0.93 & 0.41 & 0.35\nl
  12$^{*}$ & 3 44 32.62 & 32 08 36.67 & 10.14 &  9.58 &  9.36 &  9.29  &   0.56 & 0.22 & 0.07\nl
  13 & 3 44 30.85 & 32 09 54.48 &  9.80 &  9.56 &  9.41 &  9.25  &   0.24 & 0.15 & 0.16\nl
  14 & 3 44 56.18 & 32 09 11.75 & 10.90 &  9.97 &  9.48 &  8.75  &   0.93 & 0.49 & 0.73\nl
  15 & 3 44 38.73 & 32 08 39.90 & 11.09 & 10.01 &  9.51 &  9.12  &   1.08 & 0.50 & 0.39\nl
  16 & 3 44 35.00 & 32 07 35.13 & 10.70 &  9.88 &  9.52 &  9.42  &   0.82 & 0.36 & 0.10\nl
  17 & 3 44 20.98 & 32 07 38.47 & 10.01 &  9.74 &  9.67 &  9.49  &   0.27 & 0.07 & 0.18\nl
  18 & 3 44 31.52 & 32 08 44.04 & 10.56 &  9.92 &  9.68 &  9.54  &   0.64 & 0.24 & 0.14\nl
  19 & 3 44 19.12 & 32 09 30.58 & 10.07 &  9.83 &  9.73 &  9.70  &   0.24 & 0.10 & 0.03\nl
  20 & 3 44 18.01 & 32 04 56.55 & 11.93 & 10.60 &  9.73 &  9.12  &   1.33 & 0.87 & 0.61\nl
  21 & 3 44 37.82 & 32 08 02.56 & 11.62 & 10.51 &  9.75 &  9.02  &   1.11 & 0.76 & 0.73\nl
  22$^{*}$ & 3 44 32.52 & 32 08 41.48 & 10.79 & 10.09 &  9.82 &  9.76  &   0.70 & 0.27 & 0.06\nl
  23$^{*}$ & 3 44 39.21 & 32 07 33.37 & 11.26 & 10.30 &  9.90 &  9.27  &   0.96 & 0.40 & 0.63\nl
  24 & 3 44 38.41 & 32 07 33.62 & 11.19 & 10.28 &  9.91 &  9.42  &   0.91 & 0.37 & 0.49\nl
  25 & 3 44 23.96 & 32 10 59.29 & 10.62 & 10.17 & 10.06 &  9.87  &   0.45 & 0.11 & 0.19\nl
  26 & 3 44 29.76 & 32 10 38.64 & 11.85 & 10.82 & 10.16 &  9.27  &   1.03 & 0.66 & 0.89\nl
  27$^{*}$ & 3 44 21.65 & 32 10 36.93 & 12.22 & 11.05 & 10.20 &  8.74  &   1.17 & 0.85 & 1.46\nl
  28 & 3 44 42.05 & 32 08 58.40 & 11.99 & 10.92 & 10.20 &  9.56  &   1.07 & 0.72 & 0.64\nl
  29$^{*}$ & 3 44 24.24 & 32 10 18.57 & 11.09 & 10.03 & 10.29 & 10.12  &   1.06 &-0.26 & 0.17\nl
  30 & 3 44 34.78 & 32 06 31.64 & 11.43 & 10.62 & 10.35 & 10.09  &   0.81 & 0.27 & 0.26\nl
  31 & 3 44 55.68 & 32 09 16.48 & 11.74 & 10.77 & 10.39 & 10.06  &   0.97 & 0.38 & 0.33\nl
  32 & 3 44 43.52 & 32 07 40.31 & 12.05 & 10.87 & 10.43 & 10.00  &   1.18 & 0.44 & 0.43\nl
  33 & 3 44 16.42 & 32 09 54.84 & 11.26 & 10.64 & 10.45 & 10.06  &   0.62 & 0.19 & 0.39\nl
  34 & 3 44 48.70 & 32 15 25.91 & 11.26 & 10.69 & 10.53 & 10.36  &   0.57 & 0.16 & 0.17\nl
  35 & 3 44 38.52 & 32 07 59.11 & 11.88 & 10.95 & 10.54 & 10.06  &   0.93 & 0.41 & 0.48\nl
  36 & 3 44 40.20 & 32 11 32.32 & 11.63 & 10.85 & 10.56 & 10.16  &   0.78 & 0.29 & 0.40\nl
  37 & 3 44 25.57 & 32 11 30.20 & 12.23 & 11.21 & 10.59 &  9.40  &   1.02 & 0.62 & 1.19\nl
  38$^{*}$ & 3 44 22.14 & 32 05 41.90 & 12.32 & 11.20 & 10.61 &  9.95  &   1.12 & 0.59 & 0.66\nl
  39 & 3 44 25.60 & 32 12 29.49 & 11.73 & 10.96 & 10.68 & 10.40  &   0.77 & 0.28 & 0.28\nl
  40 & 3 44 33.93 & 32 08 52.49 & 11.71 & 10.94 & 10.68 & 10.41  &   0.77 & 0.26 & 0.27\nl
  41 & 3 44 28.39 & 32 07 21.12 & 11.69 & 10.93 & 10.69 & 10.41  &   0.76 & 0.24 & 0.28\nl
  42 & 3 44 26.92 & 32 04 42.91 & 11.74 & 10.98 & 10.70 & 10.27  &   0.76 & 0.28 & 0.43\nl
  43 & 3 44 32.53 & 32 08 54.78 & 11.99 & 11.09 & 10.71 & 10.34  &   0.90 & 0.38 & 0.37\nl
  44 & 3 44 34.27 & 32 10 48.07 & 12.01 & 11.15 & 10.80 & 10.49  &   0.86 & 0.35 & 0.31\nl
  45$^{*}$ & 3 44 43.78 & 32 10 28.00 & 12.15 & 11.19 & 10.81 & 10.09  &   0.96 & 0.38 & 0.72\nl
  46 & 3 44 43.43 & 32 08 15.42 & 11.60 & 11.01 & 10.82 & 10.60  &   0.59 & 0.19 & 0.22\nl
  47 & 3 44 26.60 & 32 08 19.63 & 12.79 & 11.53 & 10.84 &  9.85  &   1.26 & 0.69 & 0.99\nl
  48 & 3 44 27.86 & 32 10 51.26 & 13.77 & 11.85 & 10.86 & 10.28  &   1.92 & 0.99 & 0.58\nl
  49 & 3 44 37.27 & 32 06 09.39 & 12.00 & 11.16 & 10.89 & 10.59  &   0.84 & 0.27 & 0.30\nl
  50 & 3 44 37.44 & 32 08 58.88 & 12.24 & 11.32 & 10.89 & 10.16  &   0.92 & 0.43 & 0.73\nl
  51 & 3 44 27.79 & 32 07 30.36 & 12.05 & 11.26 & 10.97 & 10.59  &   0.79 & 0.29 & 0.38\nl
  52 & 3 44 32.74 & 32 09 14.56 & 12.23 & 11.37 & 10.99 & 10.62  &   0.86 & 0.38 & 0.37\nl
  53$^{*}$ & 3 44 33.35 & 32 09 37.83 & 12.20 & 11.39 & 11.03 & 10.31  &   0.81 & 0.36 & 0.72\nl
  54 & 3 44 39.26 & 32 09 42.57 & 12.35 & 11.43 & 11.03 & 10.65  &   0.92 & 0.40 & 0.38\nl
  55 & 3 44 23.56 & 32 06 46.01 & 12.09 & 11.32 & 11.03 & 10.77  &   0.77 & 0.29 & 0.26\nl
  56 & 3 44 17.88 & 32 12 20.46 & 12.03 & 11.34 & 11.03 & 10.79  &   0.69 & 0.31 & 0.24\nl
  57 & 3 44 21.98 & 32 12 11.31 & 12.15 & 11.40 & 11.07 & 10.69  &   0.75 & 0.33 & 0.38\nl
  58$^{*}$ & 3 44 34.85 & 32 09 51.44 & 12.01 & 11.39 & 11.08 & 11.17  &   0.62 & 0.31 &-0.09\nl
  59 & 3 44 25.46 & 32 06 16.30 & 12.78 & 11.61 & 11.09 & 10.49  &   1.17 & 0.52 & 0.60\nl
  60 & 3 44 38.45 & 32 05 03.92 & 12.39 & 11.52 & 11.10 & 10.77  &   0.87 & 0.42 & 0.33\nl
  61$^{*}$ & 3 44 19.18 & 32 07 34.68 & 12.40 & 11.58 & 11.10 & 10.41  &   0.82 & 0.48 & 0.69\nl
  62 & 3 44 22.38 & 32 12 00.39 & 12.38 & 11.55 & 11.10 & 10.39  &   0.83 & 0.45 & 0.71\nl
  63 & 3 44 36.07 & 32 15 01.57 & 12.74 & 11.60 & 11.11 & 10.69  &   1.14 & 0.49 & 0.42\nl
  64 & 3 44 44.53 & 32 08 10.55 & 12.91 & 11.82 & 11.12 & 10.18  &   1.09 & 0.70 & 0.94\nl
  65 & 3 44 11.13 & 32 06 12.26 & 12.36 & 11.43 & 11.16 & 11.00  &   0.93 & 0.27 & 0.16\nl
  66 & 3 44 14.13 & 32 10 28.18 & 12.34 & 11.55 & 11.19 & 10.75  &   0.79 & 0.36 & 0.44\nl
  67 & 3 44 38.64 & 32 08 54.61 & 12.31 & 11.53 & 11.22 & 11.05  &   0.78 & 0.31 & 0.17\nl
  68 & 3 44 40.26 & 32 14 26.01 & 12.44 & 11.66 & 11.23 & 10.89  &   0.78 & 0.43 & 0.34\nl
  69$^{*}$ & 3 44 37.51 & 32 12 22.58 & 12.90 & 11.83 & 11.24 & 10.49  &   1.07 & 0.59 & 0.75\nl
  70 & 3 44 45.15 & 32 13 34.12 & 12.42 & 11.66 & 11.26 & 10.77  &   0.76 & 0.40 & 0.49\nl
  71 & 3 44 37.14 & 32 09 14.00 & 12.36 & 11.53 & 11.26 & 11.16  &   0.83 & 0.27 & 0.10\nl
  72 & 3 44 29.93 & 32 09 19.83 & 13.40 & 11.95 & 11.27 & 10.61  &   1.45 & 0.68 & 0.66\nl
  73 & 3 44 21.54 & 32 10 16.64 & 12.30 & 11.51 & 11.28 & 10.88  &   0.79 & 0.23 & 0.40\nl
  74 & 3 44 21.12 & 32 05 01.76 & 12.54 & 11.65 & 11.29 & 10.99  &   0.89 & 0.36 & 0.30\nl
  75 & 3 44 22.98 & 32 11 56.95 & 12.41 & 11.62 & 11.30 & 11.14  &   0.79 & 0.32 & 0.16\nl
  76 & 3 44 21.55 & 32 06 24.23 & 12.43 & 11.61 & 11.33 & 10.96  &   0.82 & 0.28 & 0.37\nl
  77 & 3 44 20.11 & 32 08 55.82 & 12.54 & 11.75 & 11.36 & 10.86  &   0.79 & 0.39 & 0.50\nl
  78 & 3 44 21.28 & 32 11 56.05 & 12.54 & 11.70 & 11.36 & 10.84  &   0.84 & 0.34 & 0.52\nl
  79 & 3 44 27.34 & 32 14 20.19 & 12.78 & 11.93 & 11.39 & 10.78  &   0.85 & 0.54 & 0.61\nl
  80 & 3 44 41.81 & 32 12 00.62 & 13.02 & 11.95 & 11.39 & 10.88  &   1.07 & 0.56 & 0.51\nl
  81 & 3 44 45.27 & 32 14 10.81 & 13.31 & 12.10 & 11.46 & 11.10  &   1.21 & 0.64 & 0.36\nl
  82$^{*}$ & 3 44 25.34 & 32 10 11.81 & 12.56 & 11.86 & 11.47 & 10.96  &   0.70 & 0.39 & 0.51\nl
  83$^{*}$ & 3 44 30.47 & 32 06 28.00 & 12.68 & 11.81 & 11.48 & 11.11  &   0.87 & 0.33 & 0.37\nl
  84 & 3 44 16.32 & 32 05 32.35 & 13.24 & 12.03 & 11.49 & 11.15  &   1.21 & 0.54 & 0.34\nl
  85 & 3 44 38.53 & 32 12 57.82 & 12.59 & 11.82 & 11.50 & 11.19  &   0.77 & 0.32 & 0.31\nl
  86 & 3 44 41.38 & 32 10 22.94 & 12.48 & 11.78 & 11.51 & 11.11  &   0.70 & 0.27 & 0.40\nl
  87 & 3 44 42.59 & 32 06 16.77 & 12.43 & 11.69 & 11.52 & 11.30  &   0.74 & 0.17 & 0.22\nl
  88 & 3 44 36.93 & 32 08 32.41 & 13.02 & 12.10 & 11.56 & 10.92  &   0.92 & 0.54 & 0.64\nl
  89 & 3 44 34.88 & 32 11 16.43 & 12.70 & 11.94 & 11.60 & 11.18  &   0.76 & 0.34 & 0.42\nl
  90 & 3 44 42.74 & 32 08 31.56 & 12.95 & 12.10 & 11.60 & 10.66  &   0.85 & 0.50 & 0.94\nl
  91 & 3 44 37.91 & 32 12 16.55 & 12.78 & 12.02 & 11.61 & 11.26  &   0.76 & 0.41 & 0.35\nl
  92 & 3 44 18.59 & 32 12 53.32 & 13.65 & 12.66 & 11.65 & 10.32  &   0.99 & 1.01 & 1.33\nl
  93 & 3 44 40.15 & 32 09 10.62 & 13.13 & 12.18 & 11.66 & 11.19  &   0.95 & 0.52 & 0.47\nl
  94 & 3 44 47.70 & 32 10 52.96 & 13.35 & 12.23 & 11.68 & 11.10  &   1.12 & 0.55 & 0.58\nl
  95 & 3 44 35.43 & 32 08 54.40 & 13.08 & 12.27 & 11.77 & 11.12  &   0.81 & 0.50 & 0.65\nl
  96 & 3 44 42.57 & 32 10 00.07 & 13.45 & 12.38 & 11.78 & 10.83  &   1.07 & 0.60 & 0.95\nl
  97 & 3 44 41.17 & 32 10 07.76 & 13.56 & 12.37 & 11.79 & 10.94  &   1.19 & 0.58 & 0.85\nl
  98 & 3 44 31.37 & 32 10 45.75 & 13.31 & 12.35 & 11.81 & 11.12  &   0.96 & 0.54 & 0.69\nl
  99 & 3 44 17.61 & 32 04 47.27 & 13.01 & 12.26 & 11.81 & 11.36  &   0.75 & 0.45 & 0.45\nl
 100 & 3 44 28.48 & 32 11 21.62 & 14.40 & 12.66 & 11.83 & 11.13  &   1.74 & 0.83 & 0.70\nl
 101 & 3 44 44.90 & 32 11 03.23 & 13.00 & 12.15 & 11.85 & 11.42  &   0.85 & 0.30 & 0.43\nl
 102 & 3 44 48.96 & 32 13 19.49 & 12.93 & 12.19 & 11.88 & 11.83  &   0.74 & 0.31 & 0.05\nl
 103 & 3 44 21.79 & 32 12 31.22 & 12.98 & 12.25 & 11.92 & 11.41  &   0.73 & 0.33 & 0.51\nl
 104$^{*}$ & 3 44 18.13 & 32 09 58.84 & 13.14 & 12.34 & 11.93 & 11.26  &   0.80 & 0.41 & 0.67\nl
 105 & 3 44 53.72 & 32 06 48.70 & 13.22 & 12.34 & 11.94 & 11.54  &   0.88 & 0.40 & 0.40\nl
 106 & 3 44 21.65 & 32 15 09.53 & 12.93 & 12.27 & 11.96 & 11.54  &   0.66 & 0.31 & 0.42\nl
 107 & 3 44 46.40 & 32 11 14.24 & 13.89 & 12.57 & 11.98 & 11.74  &   1.32 & 0.59 & 0.24\nl
\enddata
\tablenotetext{1}{Sources marked with a * were found to be variable by more than
0.15 magnitudes at {\it H}-band.}
\tablenotetext{2}{{\it JHK} magnitudes were taken from Lada \& Lada (1995).}
\end{deluxetable}

%% file: Haisch.tab2.tex
\begin{deluxetable}{cccccc}
\tablecaption{IR Excess Fraction vs. Spectral Type \label{spectable}}
\tablewidth{0pt}
\tablehead{Spectral Type(s)\tablenotemark{1} & {\it N$_{Region}$}\tablenotemark{2} & { \it N$_{Excess}$}\tablenotemark{3} & { \it JHKL$_{Excess}$} (\%)\tablenotemark{4}}
\startdata
OBAF & 9 & 0 & 0\nl
G & 6 & 3 & 50\nl
K & 23 & 12 & 52\nl
M & 42 & 28 & 67\nl
All & 80 & 43 & 54\nl
\enddata
\tablenotetext{1}{Spectral Types taken from Luhman et al. (1998).}
\tablenotetext{2}{Number of stars with Spectral Types within survey boundaries.}
\tablenotetext{3}{Number of stars with Spectral Types and {\it JHKL} excesses.}
\tablenotetext{4}{Counting from the M5 boundary of the reddening band and using the Cohen et al. (1981) infrared extinction law.}
\end{deluxetable}

%% file: figurelist.tex
\clearpage
\plotone{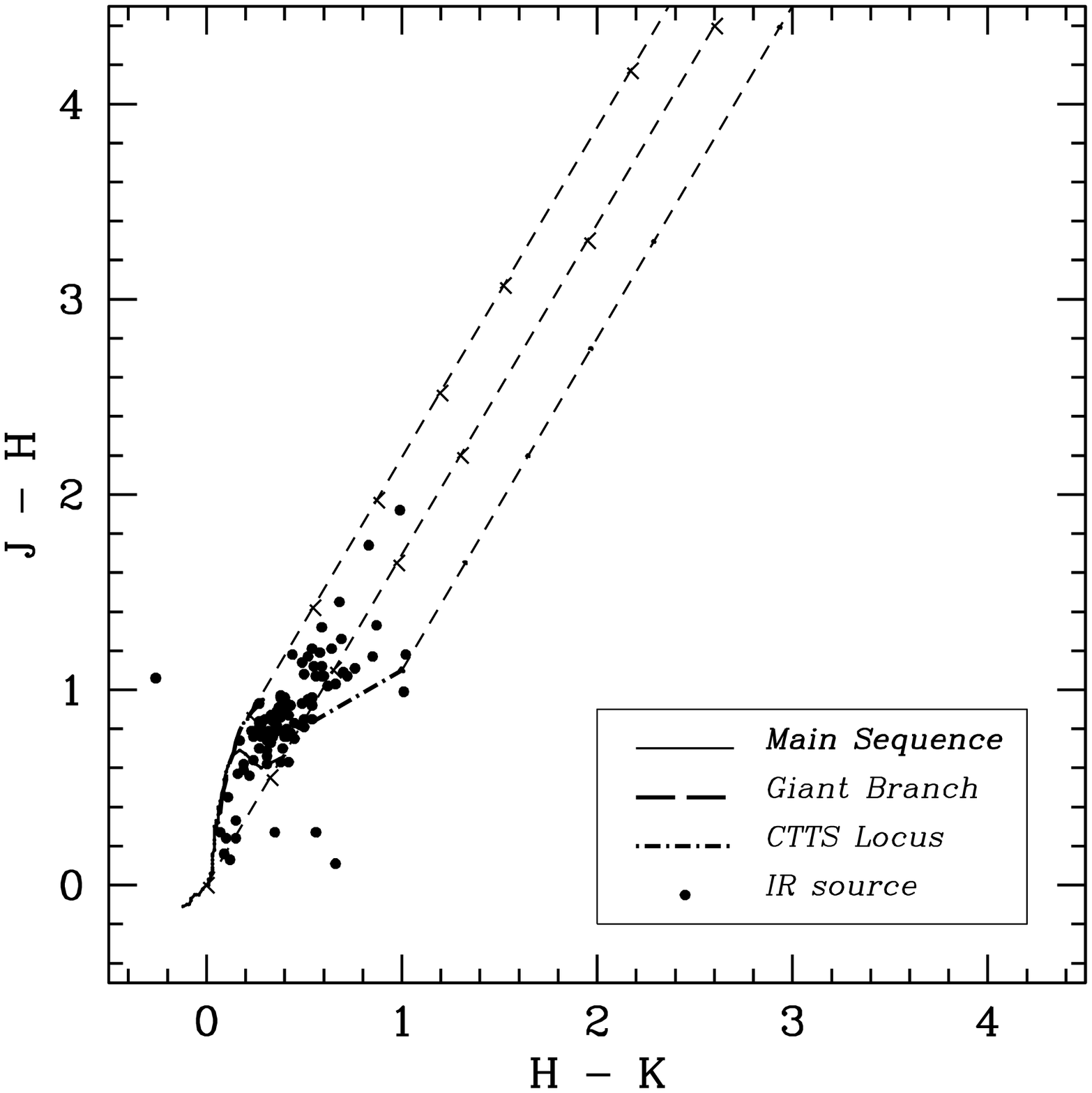}
\clearpage
\plotone{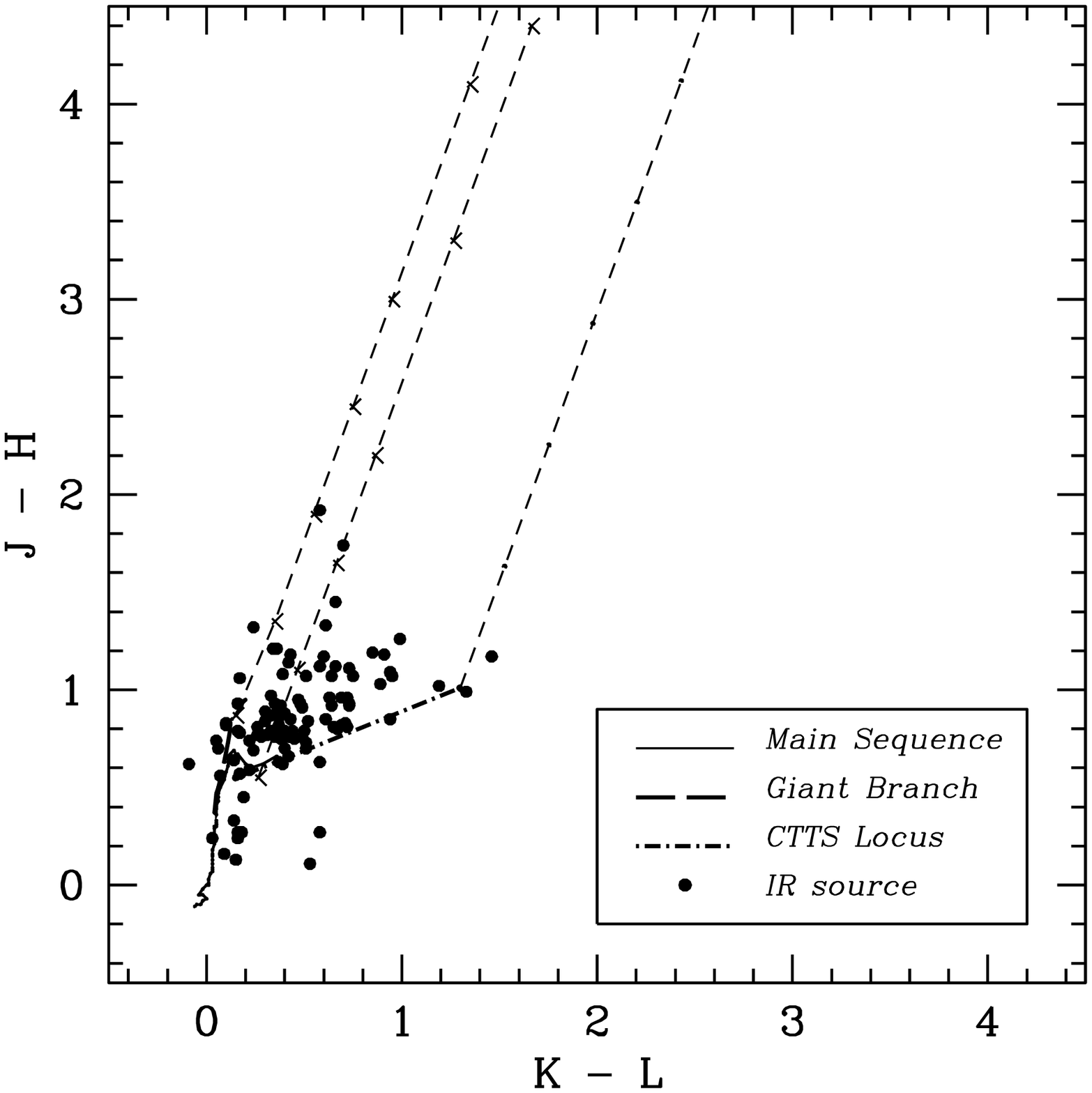}
\clearpage
\plotone{Haisch.fig2.eps}
\clearpage
\plotone{Haisch.fig3.eps}
\clearpage
\plotone{Haisch.fig4.eps}
\clearpage